\documentclass[journal]{IEEEtran}
\ifCLASSINFOpdf
   \usepackage[pdftex]{graphicx}
\else
   \usepackage[dvips]{graphicx}
\fi
\usepackage[cmex10]{amsmath}

\usepackage{array}
\usepackage{color, colortbl}

\definecolor{light-gray}{gray}{0.75}
\newcommand{\gr}{\cellcolor{light-gray}}

\usepackage{fixltx2e}

\usepackage{url}
\begin{document}
\title{Charge-Resistance Approach to Benchmarking Performance of Beyond-CMOS Information Processing Devices}
\author{Angik Sarkar, Dmitri E. Nikonov, Ian A. Young, Behtash Behin-Aein, Supriyo Datta
\thanks{A. Sarkar is currently with the Process and Device Modeling Group, Intel Corp, Hillsboro, OR 97124.  e-mail: (angik.sarkar@intel.com).}% <-this % stops a space
\thanks{D. E. Nikonov is with Components Research Group, Intel Corp, Hillsboro, OR 97124. email: (Dmitri.e.Nikonov@intel.com)}
\thanks{I. A. Young is with Components Research Group, Intel Corp, Hillsboro, OR 97124. email: (ian.young@intel.com)}
\thanks{B.Behin-Aein is currently with Technology Research Group, GLOBALFOUNDRIES USA, Sunnyvale, CA 94085, USA. email:(Behtash.Behin-Aein@globalfoundries.com)}% <-this % stops a space
\thanks{S. Datta is with Department of Electrical
and Computer Engineering, Purdue University, West Lafayette, IN 47907 USA. e-mail: (datta@purdue.edu)}
\thanks{Most of this work was done when A. Sarkar and B. Behin-Aein were with Department of Electrical
and Computer Engineering, Purdue University, West Lafayette, IN 47907 USA.}
\thanks{Manuscript received .}}

\maketitle
\begin{abstract}
Multiple beyond-CMOS information processing devices are presently under active research and require methods of benchmarking them. A new approach for calculating the performance metric, energy-delay product, of such devices is proposed.
The approach involves estimating the device properties of resistance and 
switching charge, rather than dynamic evolution
characteristics, such as switching energy and time.
Application of this approach to a wide class of charge-based and non-charge-based devices is discussed. The approach suggests pathways for improving the performance of
`beyond-CMOS' devices and a new realistic limit for energy-delay product in terms of the Plank's constant.
\end{abstract}


\section{Introduction}
\IEEEPARstart{S}{caling}  of complementary metal-oxide-semiconductor (CMOS) transistors has enabled the impressive improvement in computing power over the last few decades. However, this shrinking
of device dimensions is expected to reach a limit in the near future~\cite{Bernstein2010}. This has led to an intensive
search for the next information processing device that could potentially supplement CMOS~\cite{theis2010s}. A
wide range of physics has been invoked for different `beyond-CMOS' device proposals (see for example ~\cite{krishnamohan2008double,boucart2007double,salahuddin2008use,fiori2009ultralow,yoon2009phase,xiong2011low,banerjee2009bilayer,fiori2009possibility,chen2009dielectric,lin2010100,allwood2002submicrometer,allwood2005magnetic,imre2006majority,carlton2008simulation,datta1990electronic,appelbaum2007transit,sugahara2004spin,ren2010true,wang2005programmable,lee2007magneto,xu2008all,dery2007spin,ney2003programmable,behin2010proposal,nikonov2010proposal,khitun2008spin,khitun2009magnetoelectric}), making comparison of their performance a challenging task. In this paper, we present a simple, yet 
universal equation relating the dynamic evolution characteristics of switching energy (E) and time
($\tau$) to the device properties of resistance (R) and the switching charge (Q). This relation allows us
to propose a new approach for benchmarking and comparing the energy-delay performance of all information processing devices. 

Energy for switching all information processing devices, be it CMOS or proposed beyond-CMOS devices, has to be provided by an electrical power supply that causes charges to move through metals, semiconductors or insulators. In this paper, we use this observation to formulate a new approach for calculating the 
energy-delay performance metric of devices and circuits in terms of two quantities: 
(a) the total charge, $Q$, that is involved in switching a device;
(b) the resistance, $R$, of a device along the current path used for switching. 
$Q$, $R$ for all devices switched by a constant voltage supply, are related to the switching energy $E$ and switching delay $\tau$ 
through the central equation of this paper:
\begin{equation}
\label{etau}
E\tau=Q^2R.
\end{equation}
From a physical standpoint, it is surprising that the dynamic evolution quantity of $E*\tau$ can be determined solely from the static charge transport characteristics of Q,R even for non-electronic devices. The analytic proof of Eqn 1, valid irrespective of the state variable, is presented in Appendix  \ref{QRmetric}. In this paper, we would concentrate on the insight that can be obtained by evaluating device performance from its $Q$, $R$; since these parameters can be related to the underlying physical mechanisms of device switching along with pertinent assumptions, and fundamental and technological limitations. This `$Q-R$ approach' essentially reduces the `search for next logic switch'~\cite{theis2010s} to a problem of identifying the device physics that would allow operation with low Q and low R. 

\begin{figure*}[!t]
	\centering
		\includegraphics[scale=0.35]{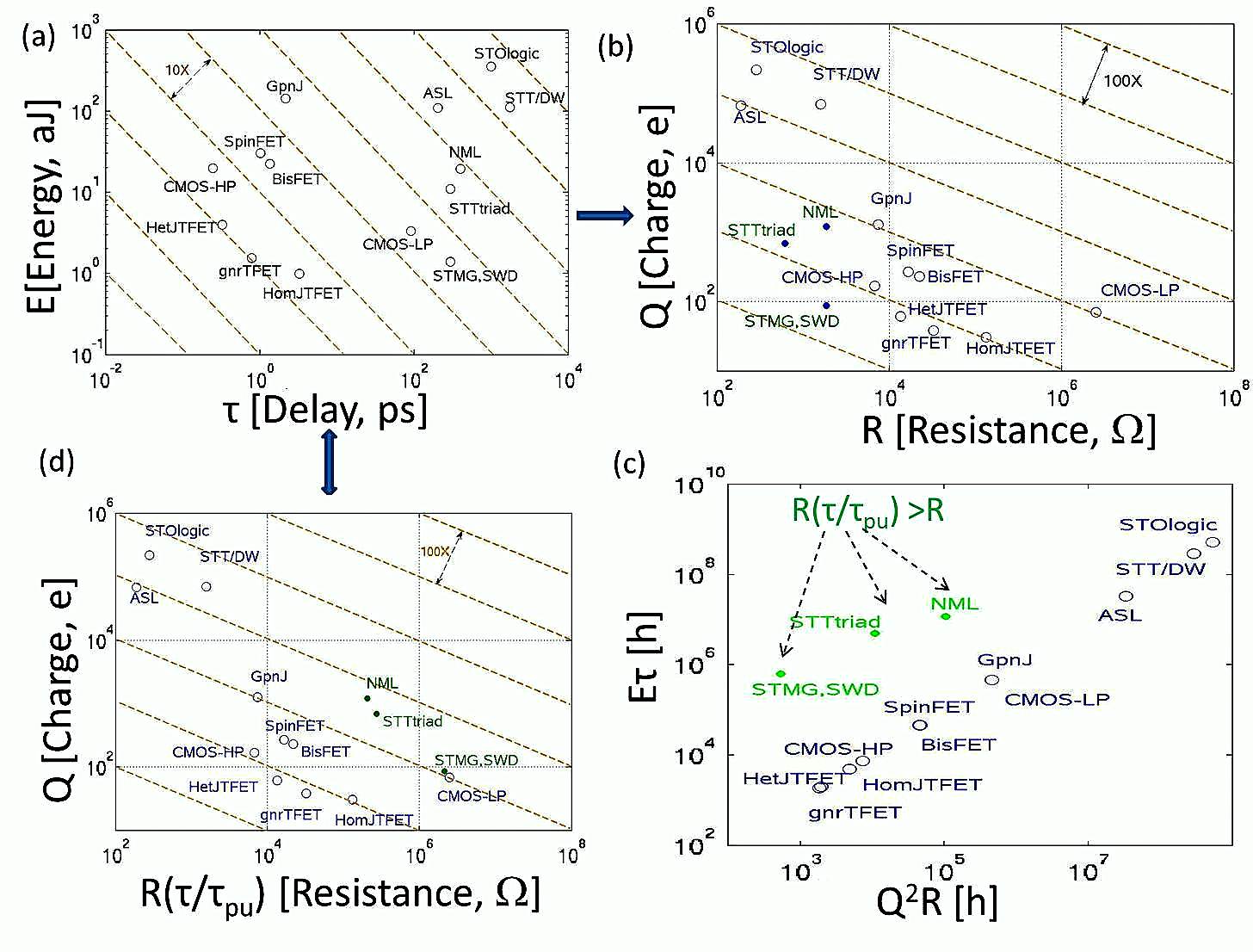}
	\caption{Comparison of device performance with the $Q-R$ approach (a) The energy-delay ($E-\tau$) 
	plot comparing the performance of various logic devices 
	(data taken from Ref~\cite{Nikonov2012}) (b) $E-\tau$ can be translated 
	to a charge-resistance ($Q-R$) plot using Eq. \ref{etau} that gives more 
	insight into device performance. 
	(c) However Eq. \ref{etau} is not valid for devices where $\tau>\tau_{pu}$, 
	the pulse width (d) To put all devices on the same footing it might be better 
	to translate the $E*\tau$ plot to an analogous $Q$-$R(\tau/\tau_{pu})$ plot. 
	Similar to the equi $E*\tau$ lines in (a), we draw equi $Q^2R$ (b) and equi $Q^2R(\tau/\tau_{pu})$ 
	lines in (d) to enable comparison between different devices. 
	Please refer to Appendix. Table I for data and device nomenclature.The acronyms used are as follows: CMOS-HP (high performance CMOS), CMOS-LP (low power CMOS), HomJTFET (Homojuntion III-V tunneling FET), HetJTFET(Heterojunction TFET), gnrTFET (Graphene nanoribbon TFET), GpnJ (Graphene pn-junction), BisFET (Bilayer pseudospin FET), SpinFET (Sughara-Tanaka SpinFET), STT/DW (Spin torque domain wall), STMG (Spin torque majority gate), STT-triad (Spin transfer torque triad), STOlogic (Spin torque oscillator), ASL(All spin logic), SWD (Spin wave device), NML (Nanomagnetic logic); $e$: Charge of an electron (1.6$\times 10^{-19}$C), $h$: Plank's constant.}
	\label{QRviewpoint}
\end{figure*}

$E*\tau$ metric of information processing devices has traditionally been evaluated from the estimation or measurements of $E$ and $\tau$, which easily communicate their potential as a fast 
 or a low-power digital switch. It would be very relevant to ask why we need a different approach.
We emphasize that the $Q-R$ approach provides a more convenient way to estimate the same value of $E*\tau$ for all devices, while relating it to the unique underlying physics of each device. Given our long experience with transistors, the methods for estimation of $E*\tau$ of CMOS devices are well known. 
However, for beyond-CMOS devices, the $Q-R$ approach for estimating energy-delay perfomance offers two distinct advantages. 

Firstly, if we were to estimate $E$ and $\tau$ directly for such devices, we would need to 
perform complex high-speed measurements or develop a theory
of the device's dynamic evolution and obtain the energy and delay by solving time-dependent equations.
The $Q-R$ approach allows us to evaluate $E*\tau$ just from static characteristics 
- $Q$,$R$ which are directly connected to underlying physical properties of any information processing device.
For example, $Q$ is related to the supply voltage and the total 
capacitance of the current path for electronic devices. For spin-torque switched devices, 
$Q$ is proportional to the total number of Bohr Magnetons or elementary spins in the nanomagnet~\cite{behin2011switching}.

Secondly, the plethora of state variables and 
physical phenomena involved in the beyond-CMOS device proposals make it difficult 
to judge whether the projections of performance in terms of $E$ and $\tau$ 
are pessimistic, realistic or whether they even violate fundamental limits (which can be physical principles or technological.) 
Neverthless, specification of performance in terms of $Q$ and $R$, both being physical device parameters, 
often allows us to judge how far away these projections are from a practical commercial device. Consider, for example, a hypothetical switch that claims to have $E=0.01aJ$, 
$\tau=100ps$ for a 100mV operation. At first glance, these would be 
impressive characteristics for a potentially low-power switch. However, if the performance 
of the same device is expressed by $R=10^8\Omega$ and $Q=0.6$ electrons, it would 
be immediately clear that the $Q$ is too low and susceptible to noise,
while $R$ is unrealistically high, pointing to problems from leakage currents in other current paths.

\section{Charge-Resistance approach}

Recently, Nikonov \textit{et. al.}~\cite{Nikonov2012} presented a comparative study of beyond-CMOS 
devices and circuits. They compared the projected performance of various devices based on their 
position on an $E-\tau$ plot (Fig \ref{QRviewpoint}a). Ideally, we would like devices which are both low-power and low-delay and occupy the lower left corner of the $E-\tau$ plot. However, from Fig \ref{QRviewpoint}a, it is apparent that few `beyond-CMOS' devices have projected $E*\tau$ better than CMOS. The extraordinary $E-\tau$ of CMOS has been achieved by years of scaling, also evident from Moore's law~\cite{moore1998cramming}. The possiblity of scaling the beyond-CMOS devices to lower their $E*\tau$ product cannot be inferred from the $E-\tau$ plot.

Nevertheless, translating the $E-\tau$ data in Fig \ref{QRviewpoint}a into $Q-R$ (Appendix \ref{QRappl}) allows us to analyze and compare performance of devices on a $Q-R$ plot (Fig  \ref{QRviewpoint}b), which immediately suggests the scaling potential and pathways for improvement of different devices.  Similar to the $E-\tau$ plot, we would like to have devices occupying the lower left corner of a $Q-R$ plot i.e. devices which have a low Q as well as low R. However, it can be observed that in general, the electronic devices have higher R and lower Q, ranging from tens to hundreds of electrons. It may be difficult to scale the already low Q further. However, in general, Q of electronic devices can be decreased by lowering the supply voltage ($V_{DD}$) or by using a more efficient device which would allow switching with a smaller capacitance since switching of electronic devices involves charging/discharging a capacitance (C) by $V_{DD}$ such that $Q=CV_{DD}$. The R of these electronic devices may be decreased by choosing a different material, preferably metallic conductors. 

Most spintronic/magnetic devices are metallic, leading to favorable low R (Fig \ref{QRviewpoint}b); but switch tens of thousands of electrons. Thus, to make $E*\tau$ comparable or better than CMOS, Q should be reduced. Relationship of Q to the device hardware and lay-out for electronic devices is easy to comprehend (Appendix \ref{QRappl}). Since E is provided by electrons driven by an electrical power supply, Q can indeed be related to physical device parameters for most information processing devices, even for spintronic and magnetic devices. Consider for example, the case of All spin logic (ASL) device~\cite{behin2010proposal} which involves switching of nanomagnets by the phenomemon of spin transfer torque, where each injected spin-polarized electron flips one Bohr Magneton or spin in the magnet. This leads to a simple relation between Q and $N_s$, the number or spins comprising the magnet~\cite{behin2011switching}:
\[
Q\sim\frac{2qN_s}{\eta},
\]
where $\eta$ is the efficiency related to the spin to charge current ratio. Owing to large $N_s$, the Q for the state of the art magnets is much greater than corresponding electronic devices (Supp II). Thus, Q can be scaled by reducing $N_s$ in a way that does not compromise the stability of the magnet~\cite{behin2011switching}. Another way of decreasing Q is to increase $\eta$, which seems possible in the light of the recent experiments which switch huge magnets using spin Hall effect ~\cite{RalphSHE} where $\eta>1$. 

The $Q-R$ approach, applied to ASL or similar spin torque based logic devices, helped us identify the device properties that should be changed to improve the $E-\tau$ performance of these devices. A similar exercise for relating the Q,R to physical device parameters can be done for almost all information processing devices (Appendix. Table II). Q,R are sufficient to determine $E*\tau$ of any device switching with a constant $V_{DD}$. Comparison of the position of devices in the $E-\tau$ (Fig \ref{QRviewpoint}a) and $Q-R$ plot(Fig \ref{QRviewpoint}b) brings to light a few `special cases'. Some of the spintronic devices (colored in green, Fig \ref{QRviewpoint}b) appear to have low Q and low R despite not comparing favorably with other logic devices on the $E-\tau$ plot, since $E\tau>Q^2R$ for these devices(Fig \ref{QRviewpoint}c). Physically, this is because these devices are switched by a short voltage pulse($\tau_{pu}$) leading to a small Q; but the long magnet switching time ($\tau>\tau_{pu}$) affects its $E-\tau$ performance. Eqn \ref{etau} is still valid for the duration the voltage pulse is on ($t=\tau_{pu}$) and to consider such devices where $\tau>\tau_{pu}$, Eqn \ref{etau} can be rewritten as (Appendix \ref{QRmetric}):
\begin{equation}
\label{etautaupu}
E\tau=Q^2R\left(\tau/\tau_{pu}\right) ; \tau\geq\tau_{pu}
\end{equation}
Eq. \ref{etautaupu} is universally valid for all information processing devices switched 
by a voltage pulse of constant amplitude. Thus, to compare all devices on the same footing irrespective of the relation between $\tau$ and $\tau_{pu}$, it might be better to compare the plot of Q vs R($\tau/\tau_{pu}$)(Fig \ref{QRviewpoint}d) which is analogous to the $E-\tau$ plot for all logic devices. Indeed in Fig \ref{QRviewpoint}d, the position of the special cases outlined in Fig \ref{QRviewpoint}c reflect their position on the $E-\tau$ plot. 

\section{Limit of Energy-Delay}

One of the main advantages of re-mapping $E-\tau$ to $Q-R$ is the evaluation of the possibility of scaling down $E*\tau$ product by scaling Q,R. Thus, it is important to investigate the minimum possible $E*\tau$ product for information processing devices. $E*\tau$ product involved in the switching a single electron is limited by the Uncertainty relation~\cite{hilgevoord1998uncertainty}:
\[
E\tau\geq h
\]
which relates $E*\tau$ to $h$, the Plank's constant. While the uncertainty relation is definitely valid for a multi-electron switching event, it may be possible to have a stricter lower bound on the $E*\tau$ product depending on whether the electrons are switching simultaneously, sequentially or a combination of both. Such a limit for $E*\tau$ product, involved in any fast switching event, can be presented using Eqn \ref{etau} since Q is fundamentally quantized and R is limited by the interface resistances.  

Consider a switching event which involves N charges:
\[Q=Nq.\]
If the charges pass through a M-moded resistor, the resistance of a device is limited by
\[
R\geq\frac{h}{q^2\sigma M}
\]
where $\sigma$ is the effective `spin' of the electrons ($\sigma=2$ if we consider the up and down spin of electrons, $\sigma=4$ for graphene due to its `pseudospin'~\cite{geim2007rise}). 
If we apply these values of Q,R to Eqn \ref{etau} we get an important result:
\begin{equation}
Q^2R=E\tau\geq h\left(\frac{N^2}{\sigma M}\right)
\label{etaulimit}
\end{equation}
where each of the three quantities (N, $\sigma$,M) on the right side of Eqn \ref{etaulimit} are well-defined physical quanties.

It is easy to see that if we consider charging a capacitor with a single electron through a single moded conductor, Eqn \ref{etaulimit} reduces to the uncertainty relation. However, most real devices involve charging of at least tens to hundreds of electrons and Eqn \ref{etaulimit} predicts a realistic stricter lower bound on $E*\tau$ product than that given by the uncertainty relation for these devices. An explanation of the limit in Eqn \ref{etaulimit} can be suggested if we consider the switching event as a transport of N electrons through M parallel paths or modes in (N/$\sigma M$) sequential batches.

Feynman~\cite{Feynman2000} argued that if we arrange \textit{sp} logic units: \textit{p} parallel rows of \textit{s} logic units in series, the net energy-delay product would be \[E\tau=ps^2E_0\tau_0,\] where $E_0$ and $\tau_0$ are energy and delay for each unit. The transport of N electrons can be considered as N logic events of switching one electron, $\sigma M$ of which occur simultaneously through the M modes. If $E_0$-$\tau_0$ is the energy-delay involved in the switching of one electron, the transport of N electrons, sequentially in (N/$\sigma M$) batches, through $\sigma M$ parallel paths should have an energy delay of \[E\tau=\sigma M(N/\sigma M)^2E_0\tau_0\],since $p=\sigma M$ and $s=(N/\sigma M)$ in this case. Assuming, $E_0$ and $\tau_0$ abide by the uncertainty principle, we get the relation in Eqn \ref{etaulimit} for the case of switching N electrons. 
\begin{figure}[!t]
	\centering
		\includegraphics[scale=0.7]{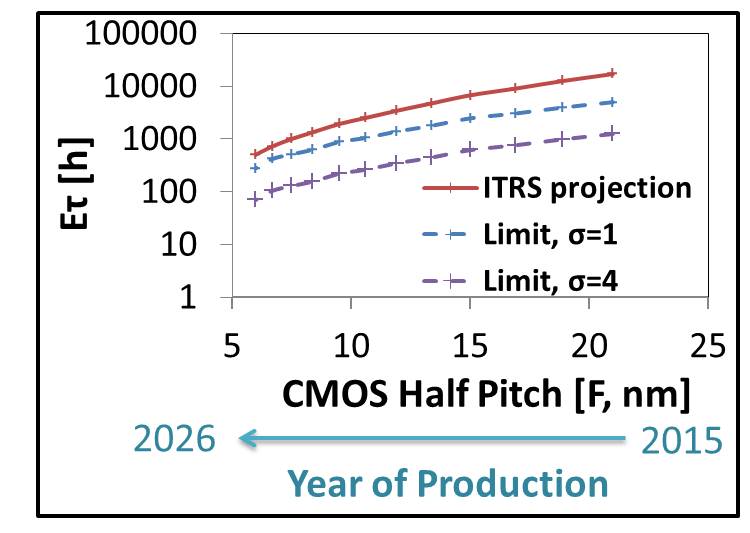}
	\caption{Comparison of the projected energy-delay of CMOS-HP to the fundamental limits of $E*\tau$ according to Eqn \ref{etaulimit}. $E*\tau$ has been calculated from the CMOS-HP projection numbers in ITRS tables~\cite{ITRS} for each technology node defined by the metal half-pitch (F). The width of the device has been assumed to be 4F. Femi wavelength($\lambda_f$) used in the calculation of the number of modes, M has been assumed to be 11nm corresponding to a carrier density of 5$\times10^{12}$/cm$^2$.}
	\label{EtauF}
\end{figure}

Eqn \ref{etaulimit} indicates that energy-delay product can possibly approach \textit{h} only for a scaled system with low N if the charges are transported through metals where M is high. Scaling of CMOS has reduced the N at the device level; we estimate N=168 for CMOS-HP projections in Ref \cite{Nikonov2012}. However, M is low in these scaled semiconductors. Considering the width (W) of 60nm \cite{Nikonov2012} and Fermi wavelength ($\lambda_f$) of $\sim$11nm, M can be estmated to be $\sim$11 using $M=Int[W/(\lambda_f/2)]$~\cite{Datta1997}. According to Eqn \ref{etaulimit}, this leads to $E\tau\geq2465h$ for CMOS-HP. This crude estimate is ~3 times lower than the $E*\tau$ product of CMOS (Fig \ref{EtauF}), assuming $\sigma=1$. However, scaling of CMOS makes it more efficient, operating closer to the $E*\tau$ limit (Fig \ref{EtauF}). 

The above example of evaluating the limits of CMOS operation brings forth the utility of the quantum limit in Eqn \ref{etaulimit} for practical information processing devices. It is well known that there is a trade-off in switching energy and delay; faster switching requires higher overdrive and higher energy consumption. Infinitely slow or adiabatic switching requires a minimum reversible energy expenditure, sometimes referred to as the Landauer limit ~\cite{landauer1961irreversibility}. Currently, this limit is of academic interest since most future-generation CMOS and `beyond-CMOS' devices aim to operate at the other extreme of switching with as low $\tau$ as possible without raising the energy consumption of the device above present day CMOS. The limit in Eqn \ref{etaulimit} addresses this energy-delay trade-off. For a real device, Eqn \ref{etaulimit} helps us estimate how low the $E*\tau$ product can be, given the technological and cost constraints that determine N,$\sigma$ and M.  

\section{Conclusion}

In essence, the $Q-R$ approach presented in this paper helps us appreciate the insight into device characteristics that can be obtained from the energy-delay performance metric. If a device is projected to have spectacular energy-delay performance, the $Q-R$ approach immediately indicates that we need to investigate what device physics leads to low Q,R. A low projected device Q (say tens of electrons, Appendix Table I), would raise questions on the effect of various electronic noise sources and device parasitic capacitances on the value of Q. A low R (tens of Ohms) would raise questions on the effect of parasitic resistances on the projected value of R. On the other hand, if a device has unimpressive energy-delay, the $Q-R$ approach can help us judge whether the energy-delay can potentially scaled down further. Essentially, the $Q-R$ approach reduces the `search for next switch' to a problem of identifying the device physics that would allow operation with low Q and low R.

We end the paper by noting that the $Q-R$ approach presented for individual devices, can also be extended to larger systems and circuits. We have applied the $Q-R$ approach to evaluate $E*\tau$ of few simple logic circuits, where it can seamlessly incorporate even the parasitic charge and resistance of interconnects. Neglecting parasitics, the total Q of such circuits is simply the product of individual device Q and the total number of devices involved in the circuit. However, Q in the presence of interconnect charge and the relation of overall circuit R to device R is non-trivial and we leave the treatment to future publications. Nevertheless, the possibility of extending $Q-R$ approach to larger systems also suggests the validity of the limit on $E*\tau$ product for larger systems, where it could provide the quantitative formulation of the energy-delay tradeoff that is fundamental to the operation of most systems.

\appendices

\section{The Charge-Resistance Metric}

\label{QRmetric}

Our aim in this section is to justify the simple relation 
for the energy (E)-delay ($\tau$) product introduced in the paper: 
\begin{equation}
\label{etau}
E\tau=Q^2R.
\end{equation}
where  $Q$ is the total charge that is involved in switching a device; $R$ is the resistance of a device along the current path used for switching.  
The relation can be easily derived for the simplified case of constant 
voltage $V_{DD}$ and current $I$ in a current path
with capacitance $C$ and resistance $R$. Then the energy and delay are:
\[E=CV_{DD}^2=QV_{DD},\]
\[\tau=Q/I,\]
and their product is
\[E\tau=RC^2V_{DD}^2=RQ^2\] 
which is the same as Eq. \ref{etau}. 
The relation between $E-\tau$ and $Q-R$ in case of capacitor charging has been pointed out by other authors 
(see for example Ref.~\cite{Feynman2000}). However, in this section, we introduce Eq. \ref{etau} as a generic 
relation valid for all information processing devices that derive the switching energy from a voltage pulse with constant amplitude $V_{DD}$. 

Consider a generic switch in which the voltage is applied as a pulse of duration $\tau_{pu}$:
\begin{equation*}
V(t)=
\begin{cases} 
 V_{DD} & t\leq \tau_{pu} \\
 0 & t>\tau_{pu}
 \end{cases}
\end{equation*}
For most switching events (e.g. the case of switching a capacitor), the switching 
time is limited by the charging time
$\tau=\tau_{pu}$. 
However, in some logic devices, e.g. spintronic devices switched by 
magnetostrictive effect(see for example Ref ~\cite{shabadi2011spin}), switching of the state variable (magnetization) 
takes much longer than the time period over which the voltage pulse is turned on. 
However, no energy is supplied by the source in this process since V(t)=0 for  $t>\tau_{pu}$. 
Hence the energy supplied by the power supply can be evaluated by integrating 
the power supplied from t=0 to t=$\tau_{pu}$
\begin{equation}
E=\int_0^{\tau_{pu}}V(t)I(t)dt=V_{DD}Q
\label{energy}
\end{equation}
where $Q=\int_0^{\tau_{pu}}I(t)dt$. If the current varies as a function of time, 
the average current can be defined  as $I_{avg}=\int_0^{\tau_{pu}}I(t)dt/\tau_{pu}$. In other words,
\[\tau_{pu}=Q/I_{avg}\]
Thus the relations for E and $\tau_{pu}$ lead to 
\[
E\tau_{pu}=Q^2(V_{DD}/I_{avg})=Q^2R
\]
which is a generalization of Eq. \ref{etau}. 
$Q$ and $R$ can thus be related to E$\tau$ by the general relation:
\begin{equation}
E\tau=Q^2R\left(\tau/\tau_{pu}\right) ; \tau\geq\tau_{pu}
\label{etautaupu}
\end{equation}
Eq. \ref{etautaupu} is thus universally valid for all information processing devices switched 
by a voltage pulse of constant amplitude $V_{DD}$.

\begin{table*}[!t] 
\caption{The Q,R extracted from the data in Nikonov and Young's paper~\cite{Nikonov2012} for various logic devices. The devices above the horizontal line have $\tau=\tau_{pu}$ while those below it have $\tau\geq\tau_{pu}$. Devices which claim to switch $Q<100e$ and $R< 1k\Omega$ are shaded. The effect of parasitic and noise Q,R may be more pronounced in these devices. The acronyms used are as follows: CMOS-HP (Si MOSFET high performance), CMOS-LP (Si MOSFET low power), HomJTFET (Homojuntion III-V tunneling FET), HetJTFET(Heterojunction TFET), gnrTFET (Graphene nanoribbon TFET), GpnJ (Graphene pn-junction), BisFET (Bilayer pseudospin FET), SpinFET (Sughara-Tanaka SpinFET), STT/DW (Spin torque domain wall), STMG (Spin torque majority gate), STT-triad (Spin transfer torque triad), STOlogic (Spin torque oscillator), ASL(All spin logic), SWD (Spin wave device), NML (Nanomagnetic logic); $e$: Charge of an electron (1.6$\times 10^{19}$C), $h$: Plank's constant}
\label{QRtable}
\centering
\begin{tabular}{cccccccc}
\hline
Device& State  & Voltage & Delay ($\tau$) & Pulse width & $E\tau=Q^2R(\tau/\tau_{pu})$ &  Charge & Resistance\\
         &Variable& V($_{DD}$) V & ps & ($\tau_{pu}$) ps      &   h & (Q) e &   (R) $\Omega$\\\hline\hline
CMOS-HP    & Charge & 0.73 & 0.25  	    & 0.25   &7.4$\times10^3$   & 168                 &6.8$\times10^3$\\
CMOS-LP     & Charge & 0.3  & 92.08 	    & 92.08 &4.6$\times10^5$   & \gr 69           		&2.5$\times10^6$\\
HomJTFET   & Charge & 0.2  & 3.27          & 3.27     & 4.8$\times10^3$ &\gr 31           		 &1.3$\times10^5$ \\    
HetJTFET    & Charge  & 0.4   & 0.33        & 0.33      &2.0$\times10^3$ & \gr 61        		     &1.3$\times10^4$ \\
gnrTFET     & Charge   & 0.25 & 0.79        & 0.79      &1.8$\times10^3$   &\gr 38         		     &3.2$\times10^4$ \\
GpnJ          & Charge   & 0.7  & 2.17         & 2.17       &4.7$\times10^5$    &127     		     &7.4$\times10^3$ \\
BisFET        & Exciton   & 0.6  & 1.36        & 1.36   	 &4.5$\times10^4 $   &230        		  &2.2$\times10^4$ \\
SpinFET     & Charge    & 0.7   & 1.02       & 1.02  	  &4.6$\times10^4$    &269       		   &1.7$\times10^4$\\
STT/DW     & Spin        & 0.01 & 1762.90  & 1762.90 & 3.0$\times10^8$   &6.9$\times10^4$ &1.6$\times10^3$\\
STOlogic    & Spin        & 0.01   & 1000.00 & 1000.00 &5.3$\times10^8$  &2.2$\times10^5$ & \gr 284 \\
ASL           & Spin        & 0.01   & 205.16   & 205.16   & 3.4$\times10^7$ &6.8$\times10^4$ & \gr 190 \\\hline
STMG         & Spin       & 0.1     & 297.61   & 0.25       &6.2$\times10^5$  & \gr  86                      &1.8$\times10^3$\\
STT-triad    & Spin       & 0.1     & 298.03   & 0.67       &4.9 $\times10^6$  &  682                  & \gr 615\\
SWD          & Spin       & 0.1     & 297.61   & 0.25       &6.2$\times10^5$  & \gr 86                      &1.8$\times10^3$\\
NML           & Spin       & 0.1     & 400.00   & 0.36       &1.2$\times10^7$  & 1.2$\times10^3$ &1.8$\times10^3$\\\hline
\end{tabular}

\end{table*}

\section{$Q-R$ approach for beyond-CMOS devices}

\label{QRappl}

In the previous section, we discussed that Eq. \ref{etautaupu} holds if (a) energy is supplied by 
voltage pulse with constant amplitude and, (b) $\tau\geq\tau_{pu}$. These conditions are valid for 
most beyond-CMOS devices, especially those considered in Ref~\cite{Nikonov2012}. As such, the 
Q, R for different devices 
can be extracted (Table \ref{QRtable}) from the E,$\tau$,$\tau_{pu}$ data in Ref \cite{Nikonov2012} 
using Eqs. \ref{energy} and \ref{etautaupu}. In this paper we have only considered the devices 
investigated by Nikonov and Young \cite{Nikonov2012}. These device broadly span different 
state variables of charge (CMOS-HP (Si MOSFET high performance), CMOS-LP (Si MOSFET low power), 
HomJTFET (III-V tunneling FET), HetJTFET(Heterojunction TFET), gnrTFET (Graphene nanoribbon TFET), 
GpnJ (Graphene pn-junction), BisFET (Bilayer pseudospin FET), SpinFET (Sughara-Tanaka SpinFET)) 
and spin/magnetization (STT/DW (Spin torque domain wall), STTMG (Spin torque majority gate), 
STT-triad (Spin transfer torque triad), STOlogic (Spin torque oscillator), ASL(All spin logic), 
SWD (Spin wave device), NML (Nanomagnetic logic)). 

To obtain the Q,R numbers in Table \ref{QRtable}, we utilized Eqns \ref{energy},\ref{etautaupu} and the E,$\tau$ data in Ref.~\cite{Nikonov2012}. However, as mentioned earlier, the benefit of the $Q-R$ approach is Q,R can be easily estimated from device parameters. For most charge based devices, Q,R can be easily related to: Q is the charge that is stored in a capacitor involved in the switching event; while R is the resistance through which the capacitor is charged. Consider for example the case of CMOS-HP. The Q and R numbers (Table \ref{QRtable}) for CMOS-HP can be easily reconciled to by considering the numbers on the ITRS
(International Technology Roadmap for Semiconductors~\cite{ITRS}) roadmap for 15 nm MOSFETs, namely\\
Supply voltage, V = 0.73 volts\\
On Current, $I_{ON}$ = 1.805 mA/m\\
Intrinsic delay ($\tau$)=0.25ps. \\ Assuming a width of W = 4F (F=15nm) =0.06 $\mu$m, we
have\\
$Q = I_{ON}\tau\sim168e$ where e is the charge of an electron.\\ The resistance $R=V_{DD}/I_{ON}=6.7\times10^3\Omega$.\\
Hence $Q^2R$ = 0.0046fJ-ps $\sim$ 7355h\\ where h is the Plank's constant. \\Analysis of the device from the $Q-R$ viewpoint immediately suggests that to improve the performance of CMOS-HP or similar charge based devices, we should move towards devices with smaller resistance (or lower $V_{DD}$, the same or slightly higher $I_{ON}$) and attempt to decrease the charge by lowering $V_{DD}$ or by using a more efficient device which would allow switching with a smaller capacitance of the switch. 

Relationship of Q to physical device parameters for electronic devices is easy to comprehend. We mentioned in the paper that Q can be related to physical device parameters for most devices, even for spintronic and magnetic devices e.g. All spin logic (ASL) device where the information is stored in the magnetization of magnets which communicate via spin currents. The Q, in this case, is easily understood from Ref~\cite{behin2011switching} which establishes that
\begin{equation}
Q\sim\frac{2qN_s}{\eta}
\label{QNs}
\end{equation}
where $\eta$ is the efficiency related to the spin to charge current ratio, assumed
to be 0.2 based on today's devices; $N_s$ is the number of Bohr Magnetons or spins comprising the magnet and is given in general by
\[
Ns=\frac{2Ku_2\Omega}{\mu_BH_k}
\]
where $Ku_2$ is the uniaxial anisotropy constant, $\Omega$ is the volume of the magnet such that $Ku_2\Omega$ is the energy barrier for magnet switching, $\mu_B$ is the Bohr Magneton and $H_k$ is the uniaxial anisotropy field.
In Ref ~\cite{Nikonov2012}, Nikonov and Young assumed the barrier for magnet switching, $Ku2\Omega = 90 kT , H_K= 3T$ for ASL. This translates to $N_s\sim15000$. Though this value of $N_s$ corresponds to a relatively small magnet compared to those used in spin-torque experiments (see for example Ref~\cite{yang2008giant}), the corresponding Q is huge, contributing to the higher $E\tau$ of ASL and other spin torque switching based switches. 

\begin{table}[!t] 
\caption{Q, R can be related to physical device parameters, even for devices with state variable other than charge. This allows improvement of E$\tau=Q^2R$ by tuning device parameters. (C-capacitance, Nc-number of charges in the capacitor, $I_{ON}$-ON current,  $N_s$- number of Bohr Magnetons comprising a magnet, $\eta$- Injection efficiency~\cite{behin2011switching} $B_{wi}$-required magnetic induction from a wire, $p_{m}$-wire width, $t_{nm}$-emperical switching time of nanomagnets in an array, $\mu$-Permeability around the wire, $\mu_0$-Permeability of vacuum, $R_{coil}$-coil resistance, R$_{junc}$-resistance of the interface between the magnet and the channel, $I_{ms}$-current required to change the polarization of the piezoelectric material, $P_{ms}$-remnant polarization, A-area of magnet).}
\centering
\begin{tabular}{ccc}
\hline
Device&  Charge (Q) &  Resistance (R) \\\hline\hline
Electronic devices    & $CV_{DD}$ & $V_{DD}/I_{ON}$ \\
(e.g. CMOS)            &              &                             \\
         &              &                             \\
\textit{Switching of magnet with current} &&\\
Spin torque (e.g. ASL)	& $2qN_s/\eta$	& $R_{junc}$\\
Magnetic field (NML) &	$2B_{wi}p_mt_{nm}/\mu\mu_0$ &  $R_{coil}$\\
         &              &                             \\
\textit{Voltage switching of magnet}&&\\
NML, SMG, SWD, STT triad&	$P_{ms}A+CV_{DD} $&$V_{DD}/I_{ms}$\\\hline
\end{tabular}
\label{QRrelation}
\end{table}

A similar exercise for relating the Q,R to physical device parameters can be done for almost all information processing devices. Such an exercise would help identify the device properties that should be changed to improve the E$\tau$ performance. The relationship of Q and R to physical device parameters for a few logic devices has been summarized in Table \ref{QRrelation}.The expressions of Q,R can be easily derived from the discussion describing the operation of these devices presented in Ref~\cite{Nikonov2012}.

\section*{Acknowledgment}

A.S. would like to thank C. Weber, J. Appenzeller, K. Roy for helpful discussions and the Center for Science of Information (CSoI), an NSF Science and Technology Center, for supporting this study at Purdue University under grant agreement CCF-0939370.

\bibliographystyle{IEEEtr}
\bibliography{QRbiblio3}

\begin{thebibliography}{10}

\bibitem{Bernstein2010}
K.~Bernstein, R.~K. Cavin, W.~Porod, A.~Seabaugh, and J.~Welser, ``{Device and
  Architecture Outlook for Beyond CMOS Switches},'' {\em Proceedings of the
  IEEE}, vol.~98, pp.~2169--2184, Dec. 2010.

\bibitem{theis2010s}
T.~N. Theis and P.~M. Solomon, ``It’s time to reinvent the transistor!,''
  {\em Science}, vol.~327, no.~5973, pp.~1600--1601, 2010.

\bibitem{krishnamohan2008double}
T.~Krishnamohan, D.~Kim, S.~Raghunathan, and K.~Saraswat, ``Double-gate
  strained-ge heterostructure tunneling fet (tfet) with record high drive
  currents and $<<$60mv/dec subthreshold slope,'' {\em Electron Devices
  Meeting, 2008. IEDM 2008. IEEE International}, pp.~1 --3, Dec 2008.

\bibitem{boucart2007double}
K.~Boucart and A.~Ionescu, ``Double-gate tunnel fet with high-$\kappa$ gate
  dielectric,'' {\em Electron Devices, IEEE Transactions on}, vol.~54, no.~7,
  pp.~1725--1733, 2007.

\bibitem{salahuddin2008use}
S.~Salahuddin and S.~Datta, ``Use of negative capacitance to provide voltage
  amplification for low power nanoscale devices,'' {\em Nano letters}, vol.~8,
  no.~2, pp.~405--410, 2008.

\bibitem{fiori2009ultralow}
G.~Fiori and G.~Iannaccone, ``Ultralow-voltage bilayer graphene tunnel fet,''
  {\em Electron Device Letters, IEEE}, vol.~30, no.~10, pp.~1096--1098, 2009.

\bibitem{yoon2009phase}
S.~Yoon, S.~Jung, S.~Lee, Y.~Park, and B.~Yu, ``Phase-change-driven
  programmable switch for nonvolatile logic applications,'' {\em Electron
  Device Letters, IEEE}, vol.~30, no.~4, pp.~371--373, 2009.

\bibitem{xiong2011low}
F.~Xiong, A.~Liao, D.~Estrada, and E.~Pop, ``Low-power switching of
  phase-change materials with carbon nanotube electrodes,'' {\em Science},
  vol.~332, no.~6029, p.~568, 2011.

\bibitem{banerjee2009bilayer}
S.~Banerjee, L.~Register, E.~Tutuc, D.~Reddy, and A.~MacDonald, ``Bilayer
  pseudospin field-effect transistor (bisfet): a proposed new logic device,''
  {\em Electron Device Letters, IEEE}, vol.~30, no.~2, pp.~158--160, 2009.

\bibitem{fiori2009possibility}
G.~Fiori and G.~Iannaccone, ``On the possibility of tunable-gap bilayer
  graphene fet,'' {\em Electron Device Letters, IEEE}, vol.~30, no.~3,
  pp.~261--264, 2009.

\bibitem{chen2009dielectric}
F.~Chen, J.~Xia, D.~Ferry, and N.~Tao, ``Dielectric screening enhanced
  performance in graphene fet,'' {\em Nano letters}, vol.~9, no.~7,
  pp.~2571--2574, 2009.

\bibitem{lin2010100}
Y.~Lin, C.~Dimitrakopoulos, K.~Jenkins, D.~Farmer, H.~Chiu, A.~Grill, and
  P.~Avouris, ``100-ghz transistors from wafer-scale epitaxial graphene,'' {\em
  Science}, vol.~327, no.~5966, p.~662, 2010.

\bibitem{allwood2002submicrometer}
D.~Allwood, G.~Xiong, M.~Cooke, C.~Faulkner, D.~Atkinson, N.~Vernier, and
  R.~Cowburn, ``Submicrometer ferromagnetic not gate and shift register,'' {\em
  Science}, vol.~296, no.~5575, p.~2003, 2002.

\bibitem{allwood2005magnetic}
D.~Allwood, G.~Xiong, C.~Faulkner, D.~Atkinson, D.~Petit, and R.~Cowburn,
  ``Magnetic domain-wall logic,'' {\em Science}, vol.~309, no.~5741, p.~1688,
  2005.

\bibitem{imre2006majority}
A.~Imre, G.~Csaba, L.~Ji, A.~Orlov, G.~Bernstein, and W.~Porod, ``Majority
  logic gate for magnetic quantum-dot cellular automata,'' {\em Science},
  vol.~311, no.~5758, p.~205, 2006.

\bibitem{carlton2008simulation}
D.~Carlton, N.~Emley, E.~Tuchfeld, and J.~Bokor, ``Simulation studies of
  nanomagnet-based logic architecture,'' {\em Nano letters}, vol.~8, no.~12,
  pp.~4173--4178, 2008.

\bibitem{datta1990electronic}
S.~Datta and B.~Das, ``Electronic analog of the electro-optic modulator,'' {\em
  Applied Physics Letters}, vol.~56, no.~7, pp.~665--667, 1990.

\bibitem{appelbaum2007transit}
I.~Appelbaum and D.~Monsma, ``Transit-time spin field-effect transistor,'' {\em
  Applied physics letters}, vol.~90, p.~262501, 2007.

\bibitem{sugahara2004spin}
S.~Sugahara and M.~Tanaka, ``A spin metal--oxide--semiconductor field-effect
  transistor using half-metallic-ferromagnet contacts for the source and
  drain,'' {\em Applied physics letters}, vol.~84, p.~2307, 2004.

\bibitem{ren2010true}
F.~Ren and D.~Markovic, ``True energy-performance analysis of the mtj-based
  logic-in-memory architecture (1-bit full adder),'' {\em Electron Devices,
  IEEE Transactions on}, vol.~57, no.~5, pp.~1023--1028, 2010.

\bibitem{wang2005programmable}
J.~Wang, H.~Meng, and J.~Wang, ``Programmable spintronics logic device based on
  a magnetic tunnel junction element,'' {\em Journal of applied physics},
  vol.~97, p.~10D509, 2005.

\bibitem{lee2007magneto}
S.~Lee, S.~Choa, S.~Lee, and H.~Shin, ``Magneto-logic device based on a
  single-layer magnetic tunnel junction,'' {\em Electron Devices, IEEE
  Transactions on}, vol.~54, no.~8, pp.~2040--2044, 2007.

\bibitem{xu2008all}
P.~Xu, K.~Xia, C.~Gu, L.~Tang, H.~Yang, and J.~Li, ``An all-metallic logic gate
  based on current-driven domain wall motion,'' {\em Nature Nanotechnology},
  vol.~3, no.~2, pp.~97--100, 2008.

\bibitem{dery2007spin}
H.~Dery, P.~Dalal, and L.~Sham, ``Spin-based logic in semiconductors for
  reconfigurable large-scale circuits,'' {\em Nature}, vol.~447, no.~7144,
  pp.~573--576, 2007.

\bibitem{ney2003programmable}
A.~Ney, C.~Pampuch, R.~Koch, and K.~Ploog, ``Programmable computing with a
  single magnetoresistive element,'' {\em Nature}, vol.~425, no.~6957,
  pp.~485--487, 2003.

\bibitem{behin2010proposal}
B.~Behin-Aein, D.~Datta, S.~Salahuddin, and S.~Datta, ``Proposal for an
  all-spin logic device with built-in memory,'' {\em Nature nanotechnology},
  vol.~5, no.~4, pp.~266--270, 2010.

\bibitem{nikonov2010proposal}
D.~Nikonov, G.~Bourianoff, and T.~Ghani, ``Proposal of a spin torque majority
  gate logic,'' {\em Electron Device Letters, IEEE}, no.~99, pp.~1--3, 2010.

\bibitem{khitun2008spin}
A.~Khitun, M.~Bao, and K.~Wang, ``Spin wave magnetic nanofabric: A new approach
  to spin-based logic circuitry,'' {\em Magnetics, IEEE Transactions on},
  vol.~44, no.~9, pp.~2141--2152, 2008.

\bibitem{khitun2009magnetoelectric}
A.~Khitun, D.~Nikonov, and K.~Wang, ``Magnetoelectric spin wave amplifier for
  spin wave logic circuits,'' {\em Journal of Applied Physics}, vol.~106,
  no.~12, pp.~123909--123909, 2009.

\bibitem{behin2011switching}
B.~Behin-Aein, A.~Sarkar, S.~Srinivasan, and S.~Datta, ``Switching energy-delay
  of all spin logic devices,'' {\em Applied Physics Letters}, vol.~98,
  p.~123510, 2011.

\bibitem{Nikonov2012}
D.~E. Nikonov and I.~A. Young, ``{Overview of Beyond-CMOS Devices and a Uniform
  Methodology for Their Benchmarking},'' {\em Proceedings of the IEEE,to be
  published}, 2013.

\bibitem{moore1998cramming}
G.~E. Moore {\em et~al.}, ``Cramming more components onto integrated
  circuits,'' {\em Proceedings of the IEEE}, vol.~86, no.~1, pp.~82--85, 1998.

\bibitem{RalphSHE}
L.~Liu, C.-F. Pai, Y.~Li, H.~W. Tseng, D.~C. Ralph, and R.~A. Buhrman,
  ``Spin-torque switching with the giant spin hall effect of tantalum,'' {\em
  Science}, vol.~336, no.~6081, pp.~555--558, 2012.

\bibitem{hilgevoord1998uncertainty}
J.~Hilgevoord, ``The uncertainty principle for energy and time. ii,'' {\em
  American Journal of Physics}, vol.~66, p.~396, 1998.

\bibitem{geim2007rise}
A.~K. Geim and K.~S. Novoselov, ``The rise of graphene,'' {\em Nature
  materials}, vol.~6, no.~3, pp.~183--191, 2007.

\bibitem{Feynman2000}
R.~P. Feynman and A.~Hey, {\em {Feynman Lectures On Computation}}.
\newblock Westview Press, 2000.

\bibitem{ITRS}
ITRS, ``{ITRS 2012 Executive Summary,
  www.itrs.net/Links/2012ITRS/Home2012.htm}.''

\bibitem{Datta1997}
S.~Datta, {\em {Electronic Transport in Mesoscopic Systems (Cambridge Studies
  in Semiconductor Physics and Microelectronic Engineering)}}.
\newblock Cambridge University Press, 1997.

\bibitem{landauer1961irreversibility}
R.~Landauer, ``Irreversibility and heat generation in the computing process,''
  {\em IBM journal of research and development}, vol.~5, no.~3, pp.~183--191,
  1961.

\bibitem{shabadi2011spin}
P.~Shabadi, A.~Khitun, K.~Wong, P.~K. Amiri, K.~L. Wang, and C.~A. Moritz,
  ``Spin wave functions nanofabric update,'' in {\em Nanoscale Architectures
  (NANOARCH), 2011 IEEE/ACM International Symposium on}, pp.~107--113, IEEE,
  2011.

\bibitem{yang2008giant}
T.~Yang, T.~Kimura, and Y.~Otani, ``Giant spin-accumulation signal and pure
  spin-current-induced reversible magnetization switching,'' {\em Nature
  Physics}, vol.~4, no.~11, pp.~851--854, 2008.

\end{thebibliography}

\end{document}